\documentclass[sigconf]{aamas}

\usepackage{balance} 
\usepackage{algorithm}
\usepackage{algorithmicx}%
\usepackage{algpseudocode}%
\usepackage{listings}

\usepackage{xcolor}
\usepackage{listings}

\title{ABMax: A JAX-based Agent-based Modeling Framework}

\author{Siddharth Chaturvedi}
\affiliation{
  \institution{Radboud University}
  \city{Nijmegen}
  \country{Netherlands}}
\email{siddharth.chaturvedi@donders.ru.nl}

\author{Ahmed El-Gazzar}
\affiliation{
  \institution{Radboud University}
  \city{Nijmegen}
  \country{Netherlands}}
\email{ahmed.elgazzar@donders.ru.nl}

\author{Marcel van Gerven}
\affiliation{
  \institution{Radboud University}
  \city{Nijmegen}
  \country{Netherlands}}
\email{marcel.vangerven@donders.ru.nl}

\begin{abstract}
Agent-based modeling (ABM) is a principal approach for studying complex systems. By decomposing a system into simpler, interacting agents, agent-based modeling (ABM) allows researchers to observe the emergence of complex phenomena. High-performance array computing libraries like JAX can help scale such computational models to a large number of agents by using automatic vectorization and just-in-time (JIT) compilation. One of the caveats of using JAX to achieve such scaling is that the shapes of arrays used in the computational model should remain immutable throughout the simulation. In the context of agent-based modeling (ABM), this can pose constraints on certain agent manipulation operations that require flexible data structures. A subset of which is represented by the ability to update a dynamically selected number of agents by applying distinct changes to them during a simulation. To this effect, we introduce ABMax, an ABM framework based on JAX that implements multiple just-in-time (JIT) compilable algorithms to provide this functionality. On the canonical predation model benchmark, ABMax achieves runtime performance comparable to state-of-the-art implementations. Further, we show that this functionality can also be vectorized, making it possible to run many similar agent-based models in parallel. We also present two examples in the form of a traffic-flow model and a financial market model to show the use case of ABMax.
\end{abstract}


\newcommand{\BibTeX}{\rm B\kern-.05em{\sc i\kern-.025em b}\kern-.08em\TeX}
\lstdefinestyle{codegray}{
  backgroundcolor=\color{gray!10},
  basicstyle=\ttfamily\small,
  frame=single,
  frameround=tttt,
  rulecolor=\color{gray!40},
  columns=fullflexible,
  showstringspaces=false,
  breaklines=true
}

\lstdefinelanguage{Python}{
  morekeywords={def,class,return,import,from,for,while,if,elif,else},
  sensitive=true
}
\begin{document}


\pagestyle{fancy}
\fancyhead{}


\maketitle 


\section{Introduction}
Complex systems are systems that are composed of many interacting parts. Examples of such systems include atoms in a molecule~\citep{testa2000molecules}, neurons in a brain~\citep{kulkarni2024towards}, companies in a market~\citep{foster2005simplistic}, computers on the internet~\citep{park2005internet}, or people in a population~\citep{rutter2017need}. It is important to model, simulate, and analyze such systems to uncover the properties of complex phenomena that can emerge in them~\citep{kwapien2012physical}. Examples are phase changes in matter, the emergence of intelligence in brains, the crashing of financial markets, the spread of a computer virus through the internet, or the spread of an infectious disease within a population.

One of the main approaches to model and simulate such complex systems is to use agent-based modeling (ABM). In this approach, the many interacting parts of a complex system are treated as simpler, independent agents. The merit of using this approach lies in the fact that designing these agents individually and the rules by which they interact with each other is often easier than designing the entire complex system all at once. That is, complexity emerges as a function of these simpler interactions. This technique has already been used to design complex systems in many domains such as sociology~\citep{macy2002factors}, ecology~\citep{deangelis2019decision}, economics~\citep{samanidou2007agent}, urban planning~\citep{chen2012agent}, neuroscience~\citep{ohsawa2018neuron}, biology~\citep{pio2022scaling}, and computational chemistry~\citep{fortuna2010agent}. 

With the increased availability of compute power in the form of hardware accelerators such as Graphics Processing Units (GPU) and Tensor Processing Units (TPU), it has become possible to simulate computational models of a large number of agents~\citep{xiao2019survey}. This helps to reduce the gap between these models and the actual phenomena they are trying to model. To facilitate agent-based modeling (ABM) and simulation, various software frameworks have been developed. Notable examples are the Python-based Mesa framework~\citep{python-mesa-2020}, the Julia-based Agents.jl framework~\citep{Agents.jl} and the C\texttt{++}-based Flame-GPU framework~\citep{richmond2017flame}.

In many agent-based modeling settings, being able to apply distinct updates to a dynamically chosen, variable-size subset of agents is a crucial feature for accurate ABM. This includes cases where the number of agents in the simulation is dynamic and new agents must be added with distinct initial states and parameters, or where multiple agents opt for the same resource, requiring a conflict-resolution protocol~\citep{yang2018evaluation}. In this work, we introduce ABMax, a Python-based ABM framework built on JAX~\citep{jax2018github} that provides two algorithms for applying such updates under JAX’s static-shape constraints. This capability can further be completely vectorized across homogeneous data structures, allowing the parallel simulation of multiple agent-based models.

ABMax combines ease of use with computational efficiency. The former is provided by an intuitive Python application programming interface (API), whereas the latter is provided by using JAX as the backend engine. JAX is a Python native high-performance array computing library that provides automatic vectorization across accelerators and just-in-time (JIT) compilation through simple decorators, allowing modelers to develop efficient large-scale models entirely in Python. This has been extensively showcased in the field of reinforcement learning~\citep{lu2022discovered} and computational simulations of evolution~\citep{evosax2022github}. ABMax provides the research community with these capabilities to support large-scale agent-based modeling.

The structure of this paper is as follows. Section~\ref{sec:API} introduces the ABMax API. Next, in Section~\ref{sec:sim} we describe the general design structure of an agent-based model developed using ABMax. Section~\ref{sec:implementation} describes implementational details of two algorithms, namely \texttt{Rank-Match} (RM) and \texttt{Sort-Count-Iterate} (SCI) used within ABMax to provide distinct updates to a subset of agents selected at runtime. In Section~\ref{sec:experiments}, we compare ABMax to other agent-based modeling frameworks and demonstrate scaling towards large-scale simulations. Finally, in the Discussion section, we present limitations and future work.


\section{ABMax}
\label{sec:API}
ABMax is an ABM framework based on JAX that provides a simple application programming interface (API) to design and simulate custom models. It relies on the \texttt{flax.struct} module from the Flax package~\citep{flax2020github} so that all objects remain friendly to JAX transformations like \texttt{jit} and \texttt{vmap}. 

ABMax supplies helper classes - \texttt{State}, \texttt{Params}, and \texttt{Signal}, each of which can be defined using a Python dictionary whose leaves are valid JAX types. Although these classes are interchangeable from the compiler's perspective, conceptually they serve different roles. A \texttt{State} object represents quantities that change every step, like position or energy of the agents. A \texttt{Params} object stores constants like time-step sizes or neural-network weights that remain fixed for multiple steps. \texttt{Params} is also meant to pass additional information to ABMax functions needed for execution. Objects of type \texttt{Signal} are meant to be used as messages that are passed between agents or between agents and the environment.
Building on these helpers, the framework exposes three core classes that users inherit to define a model.

First, a \texttt{Policy} represents the decision rule that agents follow to take actions in the environment. A \texttt{Policy} instance owns a \texttt{State} field for any internal memory needed by the rule, for instance, the hidden vector of a recurrent neural network (RNN), and a \texttt{Params} field to store, for instance, the RNN's connection strengths and biases. 

Second, an \texttt{Agent} encapsulates a single actor or agent in the environment. An agent bundles three parts: its mutable \texttt{State} (like position, energy, etc.), its fixed \texttt{Params} (like vision radius, metabolism rate, etc.), and the \texttt{policy} it uses to act in the environment. Further, each agent carries a unique integer \texttt{id} for unique identification during the simulation, a \texttt{type} tag that classifies the agent (species, role, or cohort), and an \texttt{age} counter that can be incremented at every step to track its lifespan.

Finally, a \texttt{Set} is a wrapper around a collection of \texttt{Agents}. It stores an array of \texttt{Agent} objects and two integers \texttt{num\_agents} and \texttt{num\_active\_agents} that represent the maximum number of agents possible and the current number of agents in the set, respectively. Further, it also mirrors most of the fields held by a single \texttt{Agent}, including a \texttt{State}, \texttt{Params}, and \texttt{Policy} object. This structure makes it easier to run batched and cascaded ABM simulations using a single \texttt{vmap}.

\begin{figure}[h]
  \centering
  \includegraphics[width=1.0\linewidth]{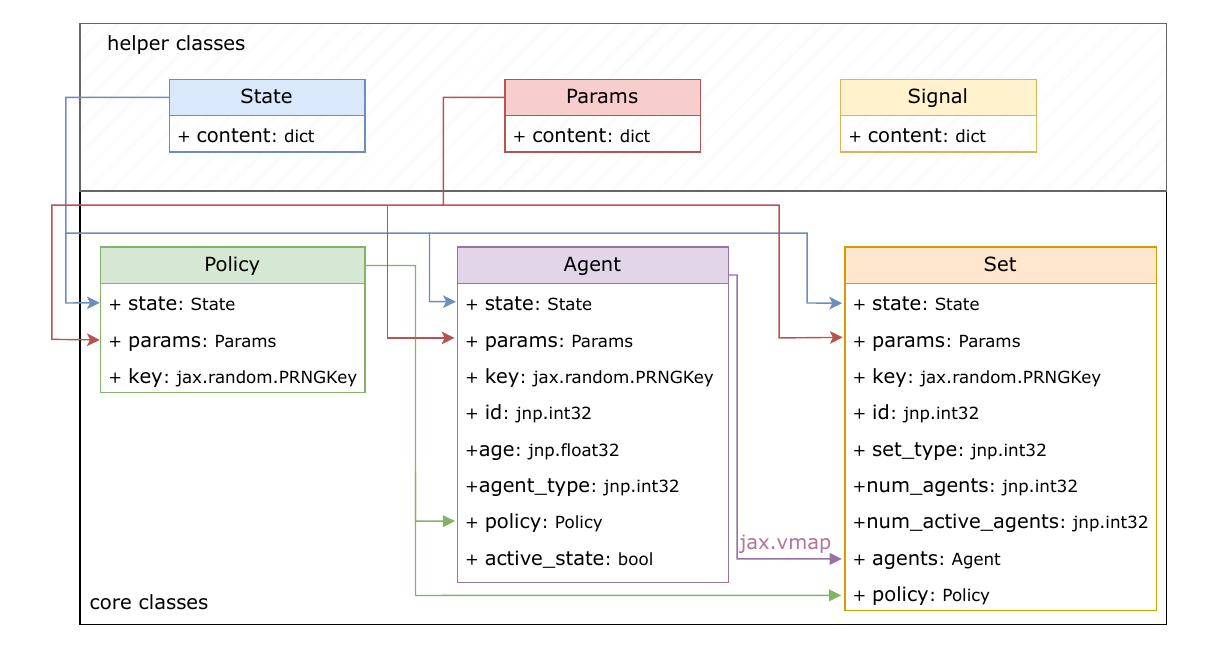}
  \caption{Unified Modeling Language (UML) diagram representing different classes in the ABMax framework. These classes can be divided into the helper and the core classes.}
  \label{fig:ABMax_uml}
\end{figure}

A unified modeling language (UML) diagram of these classes can be viewed in Figure~\ref{fig:ABMax_uml}. Once the agents have been defined, ABMax offers a concise set of functions for common ABM manipulations, including:
\begin{itemize}
    \item \texttt{create\_agents}: Instantiates a fixed-shape collection of agents by using \texttt{vmap} based on function arguments like the number of agents, the number of active agents, and a user-defined \texttt{Params} object. It initializes all the fields that are needed to define an \texttt{Agent} object.
    
    \item \texttt{step\_agents}: Applies a user-defined state-transition function to every agent in parallel. Returning an updated \texttt{Set}.
    
    \item \texttt{set\_agents\_sci}: Uses a Sort-Count-Iterate kernel as defined in Section~\ref{sec:implementation} to update $n$ agents with $n$ unique changes.
    
    \item \texttt{set\_agents\_rm}: Applies a Rank-Match kernel to $n$ selected agents and applies $n$ unique updates to them as described in Section~\ref{sec:implementation}.
    
    \item \texttt{set\_agents\_mask}: Applies independent changes to $n$ selected agents based on a mask. These changes are independent and non-conflicting, unlike the ones discussed in Section~\ref{sec:implementation}, and thus can be applied in parallel. 
    
    \item \texttt{sort\_agents}: Reorders the collection of agents either in ascending or descending order based on an ordering-key supplied during runtime. The ordering key is a one-dimensional array whose length is equal to the number of agents in the collection.

    \item \texttt{select\_agents}: Sorts the indices of the agents according to a selection condition. The indices of the agents that satisfy the selection condition are permuted to the front of the index array. Finally, the function returns this sorted list of indices and the number of agents selected.

\end{itemize}
Apart from the \texttt{create\_agents} function, all the above functions have a JIT-compiled callback. They can also be vectorized across multiple \texttt{Set} objects of agents, provided the objects have the same type and shape of leaves.


\section{Simulation Design}\label{sec:sim}
A typical design of a project based on ABMax can be seen below.

\begin{lstlisting} [style=codegray, language=Python, caption={Typical ABMax project in Python pseudo-code}]
import jax
import abmax
from flax import struct

class Agent_1(abmax.structs.Agent):
    def create_agent(...) -> abmax.structs.Agent:
    ...
    
    def step_agent(...) -> abmax.structs.Agent:
    ...
    
    def update_agent(...) -> abmax.structs.Agent:
    ...

class Agent_2(abmax.structs.Agent):
    ...

@jax.jit
def agent_interactions(...):
    ...

Class Sim():
    Agent_A1_set: abmax.structs.Set
    Agent_A2_set: abmax.structs.Set
    
    def create_sim(...) -> Sim:
    ...

    def step(...) -> Sim:
    ...

def main(...):
    sim = Sim.create_sim(...)
    sim = jax.lax.scan(Sim.step)(sim, ...)
\end{lstlisting}

Each heterogeneous agent class can be defined by inheriting the \texttt{Agent} class from ABMax. The class encapsulates a \texttt{create\_agent} function that contains the procedure to initialize a single agent; this function is later passed to \texttt{create\_agents} for instantiating a vectorized data structure of agents. Similarly, the \texttt{step\_agent} member implements the logic to step a single agent in the simulation, which is vectorized across all agents of that class by a JIT-compiled callback to \texttt{step\_agents}. The \texttt{update\_agent} member defines single-agent update logic that is applicable to a subset of agents. Depending on whether the applicability is unique or common to all agents in the subset, this member is passed as an argument to either the \texttt{set\_agents\_rm} and \texttt{set\_agents\_sci} kernels (for unique updates) or to the \texttt{set\_agents\_mask} kernel (for non-unique updates).

Next, the \texttt{agent\_interactions} function represents the functions that define how different agents interact with each other. Such functions differ across ABM contexts and thus lend themselves to a set of \emph{design patterns}. With ABMax, we aim to maintain a collection of these patterns. A separate class \texttt{Sim} can be declared to set up all the agent sets in the project and to define how the entire simulation advances one step via the members \texttt{create\_sim} and \texttt{step}, respectively. Agent-set manipulation functions like \texttt{set\_agents\_sci}, \texttt{set\_agents\_rm}, \texttt{set\_agents\_mask}, \texttt{select\_agents}, and \\ \texttt{sort\_agents} can be used in \texttt{Sim.step} or \texttt{agent\_interactions} to achieve the desired agent manipulation.

In practice, all classes are decorated with the \\ \texttt{flax.struct.dataclass} decorator so they comply with JAX transforms, and relevant class member functions are written as \\ \texttt{staticmethod}s to align with the functional programming paradigm used with JAX. Finally, \texttt{Sim} objects can be scanned using \texttt{jax.lax.scan} for the desired number of time steps to run the simulation.


\section{Implementation}\label{sec:implementation}

ABMax provides the ability to update a dynamically selected number of agents that can be determined during runtime through a JIT compiled function. This capability can also be completely vectorized across homogeneous data structures, allowing the parallel simulation of multiple models wherein a different number of agents can be updated distinctly in each of them. The main constraint of our approach is that modelers must know the maximum number of agents in advance, since we use placeholder agents to keep data structures’ shapes constant throughout the simulation. Within the ABMax framework, we present \texttt{Rank-Match} and \texttt{Sort-Count-Iterate} as two algorithms to realize this functionality, with the former focusing on speed and the latter focusing on modeling flexibility.

Use cases in which this capability is relevant can be understood by the following toy example. Consider $\mathbf{a} \in \mathbb{Z}^n$ and $\mathbf{b} \in \mathbb{Z}^m$ vectors of $n$ and $m$ random integers, respectively. Let $\mathbf{a}$ have $p\in\{1,n\}$ even integers and $\mathbf{b}$ have $q\in\{1,m\}$ odd integers. The task is to set the first $r=\operatorname{min}(p,q)$ even integers of $\mathbf{a}$ to the first $r$ odd integers of $\mathbf{b}$.
This task bears resemblance to many contexts in ABM, such as a {predation model} which requires choosing $q$ animals ready to reproduce and placing their offspring on $p$ available lattice sites, {traffic cellular-automata} where $q$ incoming cars are inserted into $p$ empty entry cells of a road or a {stock-exchange simulation} where $q$ traders' place new limit orders to the first $p$ vacant slots in the order book.
Each scenario requires applying distinct updates to or updates by a variable-sized, dynamically selected subset of agents.

\begin{algorithm}[t]
  \caption{Rank–Match (RM) algorithm}
  \label{algo:RM}
  \begin{algorithmic}[1]
    \Require
      $\mathbf{a}\in\mathbb{Z}^n$, $\mathbf{b}\in\mathbb{Z}^m,$ \quad\Comment{source integer vector $\mathbf{a}$ and target integer vector $\mathbf{b}$}\\
      $\mathbf{m_a} \gets \mathbf{a}\bmod2 == 0$, $\mathbf{m_b} \gets \mathbf{b}\bmod2 == 1$ \quad\Comment{calculate the boolean masks for $\mathbf{a}$ and $\mathbf{b}$ using \texttt{where} from JAX}\\
      $\mathbf{r_a} \gets \operatorname{cumsum}(\mathbf{m_a})\odot\mathbf{m_a}$, 
      $\mathbf{r_b} \gets \operatorname{cumsum}(\mathbf{m_b})\odot\mathbf{m_b}$
      \quad\Comment{calculate the unique ranks of $\mathbf{a}$ and $\mathbf{b}$ using \texttt{cumsum} operation from JAX}
      \For{$a, r_a \operatorname{in} \mathbf{a}, \mathbf{r_a}$} \quad\Comment{loop through $\mathbf{a}$ and $\mathbf{r_b}$ using \texttt{vmap} from JAX}
        \For{$b, r_b \operatorname{in} \mathbf{b}, \mathbf{r_b}$} \quad\Comment{loop through $\mathbf{b}$ and $\mathbf{r_b}$ using \texttt{vmap} from JAX}
        \If{$r_a == r_b$} \quad\Comment{if the ranks match}
            \State $a_c \gets b$ \quad\Comment{then apply the update (copy $b$ to a candidate of $a$)}
            \State $m_u \gets 1$ \quad\Comment{mark the update as applied}
            \State \textbf{yield} $a_c$, $m_u$
          \Else
            \State $a_c \gets a$ \quad\Comment{else, no change is applied}
            \State $m_u \gets 0$ \quad\Comment{mark the update as not applied}
            \State \textbf{yield} $a_c$, $m_u$
          \EndIf
        \EndFor
        \State collect $\mathbf{a_c}$ and $\mathbf{m}_u$\quad\Comment{collect a vector of candidate numbers $\mathbf{a_c}$ and the update mask $\mathbf{m_u}$}
        \State $i \gets \operatorname{argmax}(\mathbf{m_u})$\quad\Comment{find the index $i$ of the updated element in $\mathbf{a_c}$}
        \State $a' \gets \mathbf{a_c}[i]$ \quad\Comment{return the chosen candidate update}
        \State \textbf{yield} $a'$
      \EndFor
    \State collect $\mathbf{a}'$ \quad\Comment{collect the updated $\mathbf{a}$}\\
    \Return $\mathbf{a}'$ \quad\Comment{return the updated $\mathbf{a}$}
  \end{algorithmic}
\end{algorithm}

\textbf{Rank-Match} (RM) starts by selecting the agents that are deemed to be updated. For this, we form a Boolean mask corresponding to the agents using the \texttt{where} function from JAX. Then we calculate the cumulative sum of the mask using the \texttt{cumsum} and multiply it by the mask itself using element-wise multiplication. This assigns a zero to the agents that do not need to be updated and an ascending, unique label integer to the agents that need to be updated. We do the same procedure in case a subset of updates also needs to be selected and uniquely labeled from a larger super set. Next, we iterate through all the agents and all the changes in a vectorized way using \texttt{vmap} from JAX. During each iteration, if the unique integer label of the agent matches that of the update, we apply the update and return the updated agent along with the boolean value \textit{true}. In case the labels do not match, we return the original agent with the boolean value \textit{false}. This results in a vector of candidate agents for each agent corresponding to the number of updates, and another boolean mask where the value \textit{true} corresponds to an updated agent and \textit{false} corresponds to no change. Finally, we use the \texttt{argmax} function on this returned boolean mask to find the index of the agent from the candidate agent vector that must be returned. The application of this approach on the toy example introduced earlier can be seen in Algorithm~\ref{algo:RM}.

\textbf{Sort-Count-Iterate} (SCI) also starts by selecting the agents that are deemed to be updated. After obtaining the Boolean mask for both the selected agents and the relevant updates, we \texttt{sum} the mask elements to obtain the number of agents that need to be updated and the number of updates that must be processed. Following that, we take the minimum of the two numbers, $r$. Next, the indices of the agents and the updates are sorted using the \texttt{argsort} function from JAX, wherein the indices corresponding to which, the boolean masks read \textit{true} are permuted forward, and the indices corresponding to the value \textit{false} are shifted backwards. Finally, a shape-static while loop in the form of \texttt{lax.while\_loop} from JAX iterates through the sorted indices $r$ times during which it sequentially applies a particular update to a particular agent. 

The \texttt{Rank-Match} (RM) algorithm is more performant than the \texttt{Sort-Count-Iterate} (SCI) algorithm, as will be observed in Section~\ref{sec:experiments}. However, SCI offers additional modeling freedom by exposing and allowing explicit manipulation of agent indices in ABM. 
Algorithm~\ref{alg:SCI} presents the algorithm while solving the example task.

\begin{algorithm}[h]
  \caption{Sort-Count-Iterate (SCI) algorithm}
  \label{alg:SCI}
  \begin{algorithmic}[1]
    \Require
      $\mathbf{a}\in\mathbb{Z}^n$, $\mathbf{b}\in\mathbb{Z}^m,$ \quad\Comment{source integer vector $\mathbf{a}$ and target integer vector $\mathbf{b}$}\\
      $\mathbf{m}_a \gets \mathbf{a}\bmod2 == 0$, $\mathbf{m}_b \gets \mathbf{b}\bmod2 == 1$ \quad\Comment{calculate the boolean masks for $\mathbf{a}$ and $\mathbf{b}$ using \texttt{where} from JAX}\\
      $r \gets \operatorname{min}(\operatorname{sum}(\mathbf{m}_a),\operatorname{sum}(\mathbf{m}_b))$ \quad \Comment{calculate the number of updates using the \texttt{min} and  \texttt{sum} functions from JAX} \\
      $\mathbf{i_a} \gets \operatorname{argsort}(\mathbf{m_a})$,
      $\mathbf{i_b} \gets \operatorname{argsort}(\mathbf{m_b})$
      \quad \Comment{move the indices of the \textit{true} elements of $\mathbf{m_a}$ and $\mathbf{m_b}$ towards the front using \texttt{argsort} function from JAX}
      \For{$i = 0,\dots,r-1$}\quad\Comment{loop through the sorted indices using a \texttt{lax.while} loop from JAX}
        \State $\mathbf{a}[\mathbf{i_a}[i]] \gets \mathbf{b}[\mathbf{i_b}[i]]$
      \EndFor\\
    \Return $\mathbf{a}$ \quad\Comment{return the updated $\mathbf{a}$}
\end{algorithmic}
\end{algorithm}

\section{Experiments}
\label{sec:experiments}

We evaluated ABMax on three different tasks, that is, a predation model, a traffic simulation, and a financial market simulation. 
Certain steps of these models can be handled by the general API detailed before, but certain steps are specific to the models and require the implementation of context-informed algorithms. Thus, ABMax also aims to provide a collection of these algorithms implemented using ABMax data structures. Further, it can also be noted that, by making all the API data structures friendly to JAX transforms, ABMax enables vectorization in ABM not only at the overall model level but also at sub-model levels. Thus, different sub-parts of a model can be vectorized depending on the context. 

\subsection{Predation model}

\begin{figure}[h]
\centering
\includegraphics[width=1.0\linewidth]{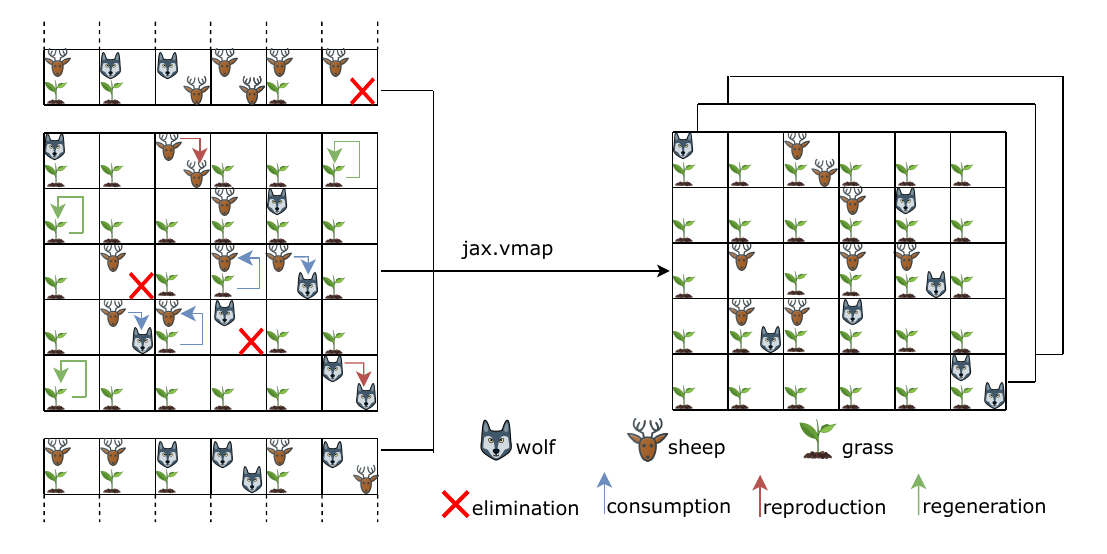}
\caption{Depiction of Wilensky’s discrete 2D predation model~\citep{Wilensky1997}. Wolves gain energy by consuming sheep, and sheep gain energy by eating grass, which regrows on each cell after a fixed delay. Both species move randomly and reproduce asexually with a given probability, spawning offspring at their current location. ABMax leverages JAX’s \texttt{vmap} to run multiple such simulations in parallel over identical data structures.
}\label{fig:ABMax_wolf_sheep}
\end{figure}

We compared the performance of ABMax against other frameworks using a canonical predation model~\citep{Wilensky1997}, in which three agent species, namely, grass, sheep, and wolves, interact on a two-dimensional grid lattice as illustrated in Figure~\ref{fig:ABMax_wolf_sheep}. 
This benchmark is a better test for the framework compared to other trivial models, for instance, the boid model~\citep{wilensky1998netlogo_flocking}, because at every step it requires distinct updates to a runtime determined subset of agents. Specifically, at each time step, a random number of sheep or wolves become ready and are selected to reproduce based on their internal energy levels. The new agents must be spawned on the same cell as the parent, making the updates unique for each agent added to the simulation. 

We compared ABMax using either RM or SCI with the Agents.jl~\citep{Agents.jl} framework developed in the Julia programming language~\citep{bezanson2012julia} and Mesa~\citep{python-mesa-2020}, a Python native ABM framework. These simulations were carried out on an Intel XEON CPU and an NVIDIA A100 GPU.
As shown in Table~\ref{tab:wolf_sheep_agents.jl}, Agents.jl does outperform ABMax because it is tuned for serial or threaded updates on dynamically sized Julia arrays, incurring virtually no padding or shape-management overhead. However, in a pure Python environment, ABMax substantially narrows the performance gap, bringing the runtime of ABM simulations much closer to that of Agents.jl than Mesa. 

\begin{table}[h]
\caption{Time taken in simulating 100 steps of the predation model reported in milliseconds for small and large environments. The small environment starts with 600 sheep and 400 wolves spawned on a 100 by 100 lattice. A large environment starts with 6000 sheep and 4000 wolves on a 1000 by 1000 lattice. The times are calculated by subtracting the median runtime of five steps from that of 105 steps over ten runs to erase the effect of initial setup time, which is substantially high for JAX. }\label{tab:wolf_sheep_agents.jl}
\centering
\begin{tabular}{l|ll}
{\bf Environment size}  & {small} & {large}\\ \midrule
{\bf Agents.jl}    &$14.93$    &$685.03$ \\
{\bf ABMax RM}    &$50.26$    &$3315.88$ \\
{\bf ABMax SCI}  &$726.78$    &$5455.01$ \\
{\bf Mesa}      &$1333.047$   &$170070.95$ \\
\end{tabular}
\end{table}

Another useful feature of ABMax is that  it can run multiple agent-based models in parallel over hardware accelerators by simply using \texttt{vmap} from JAX. 
The speed-up enabled by this feature is visible in Table~\ref{tab:wolf_sheep_vmap}. 

\begin{table}[h]
\caption{Time taken in simulating 100 steps of multiple predation models reported in milliseconds. The parameters of this model are the same as the small environment in Table~\ref{tab:wolf_sheep_agents.jl}.}\label{tab:wolf_sheep_vmap}
\centering
\begin{tabular}{ll}
number of models & time taken (ms) \\ \midrule
$10$ & $3088.58$\\
$20$ & $3171.38$\\
$50$ & $3083.56$\\
$100$ & $3316.24$\\
$200$ & $3616.16$\\
$500$ & $5343.62$
\end{tabular}
\end{table}

\begin{figure*}[h]
\centering
\includegraphics[width=0.8\linewidth]{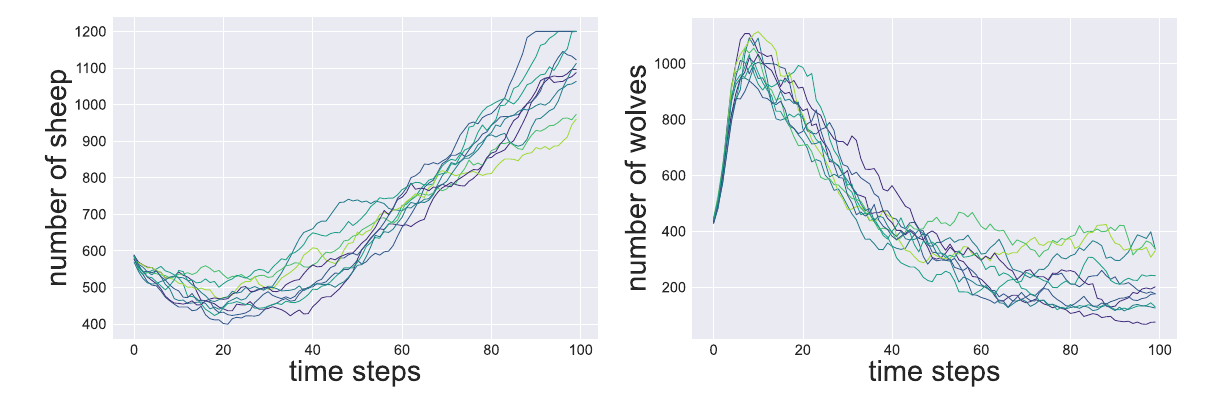}
\caption{Trend in the number of wolves and number of sheep for similar initial conditions using ten different seeds, simulated for 100 steps in a small predation environment.}
\label{fig:wolf_sheep_result}
\end{figure*}

An example of such a simulation can be seen in Figure~\ref{fig:wolf_sheep_result}, where ten populations of wolves and sheep were simulated in ten different environments, all in parallel. These simulations start with similar initial conditions but with different seeds, and thus the difference in trend reflects the stochasticity in the position and reproduction of agents.

\subsection{Traffic simulation}
\begin{figure}[h]
\centering
\includegraphics[width=1.0\linewidth]{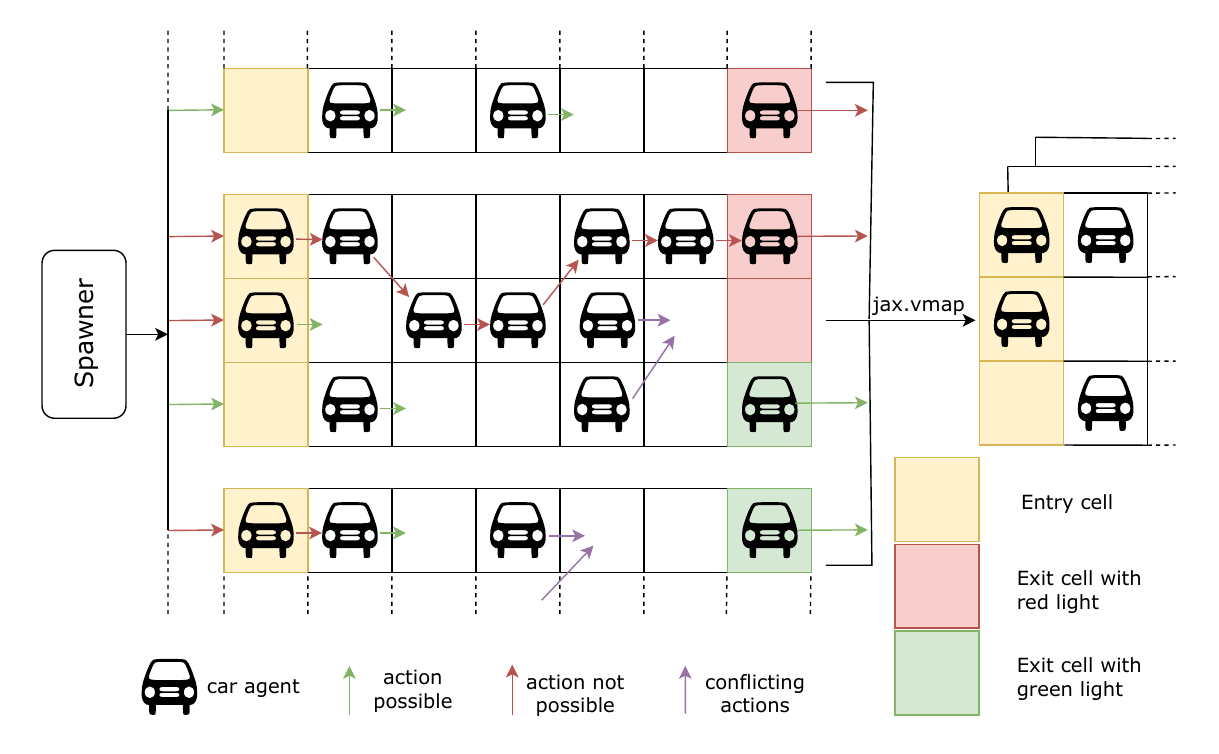}
\caption{A cellular‐automaton traffic model: green arrows mark allowed moves, red arrows blocked actions, and purple arrows potential conflicts. Cars spawn in yellow entry cells and travel toward exit cells under green/red signals. By representing everything with ABMax data structures and applying JAX’s \texttt{vmap}, we can parallelize simulations across multiple roads with different car counts.}\label{fig:ABMax_traffic}
\end{figure}

We developed a cellular-automata (CA) based simulation for traffic movement on a finite road containing three lanes as shown in Figure~\ref{fig:ABMax_traffic}. A very large-scale multi-GPU simulation of traffic flow using this approach exists in the form of LPSim~\cite{jiang2024large}. Using ABMax, we developed a light-weight version of such a model.

Compared to a continuous space model, a CA model simplifies the analysis of traffic movement and the design of rules by making traffic movement discrete. In this model, the cars are randomly introduced in the three entry cells, and they move towards the exit cells by randomly changing their lanes. It is not possible for two cars to occupy the same road cell; thus, a conflict-management algorithm is needed in the scenario. Based on a traffic-light schedule, a car can or cannot exit the road from the exit cells, which makes it possible to observe traffic jams. Once an entry cell is occupied by a car, it cannot be used to spawn new cars onto the road. This requires a car-spawning algorithm. For this model, we used ABMax data structures and functions to develop, firstly, a completely vectorized way of resolving the conflicts when multiple cars want to enter the same cell. This was done by treating each cell of the road as an agent as well and installing a priority mask with respect to cars entering the cell from neighboring cells. Secondly, a vectorized way of a random number of cars in a randomly selected subset of entry cells, while taking into account their availability, was also developed. This combination made it possible to simulate multiple roads with different numbers of active cars in parallel. 

Table~\ref{tab:traffic_tab} shows a comparison of the time taken to simulate 1000 steps of a larger model across different numbers of roads. It can be noted that when run on a GPU, the time taken to simulate a larger number of roads does not scale linearly with the number of roads themselves when the size of data structures can be efficiently accommodated on the GPU.

\begin{table}[h] 
\caption{Time taken in simulating 1000 steps of the traffic-flow simulation for different numbers of roads. The roads are three lanes wide and 100 cells long, thus capable of housing 300 cars at a time.}
\label{tab:traffic_tab}
\centering
\begin{tabular}{ll}
\text{number of roads}& \text{time taken (ms)}\\ \midrule
$10$ & $931.65$\\
$20$ & $996.60$\\
$50$ & $1028.15$\\
$100$ & $1230.85$\\
$200$ & $1467.89$\\
$500$ & $2327.71$\\

\end{tabular}
\end{table}

\begin{figure*}[h]
\centering
\includegraphics[width=0.8\linewidth]{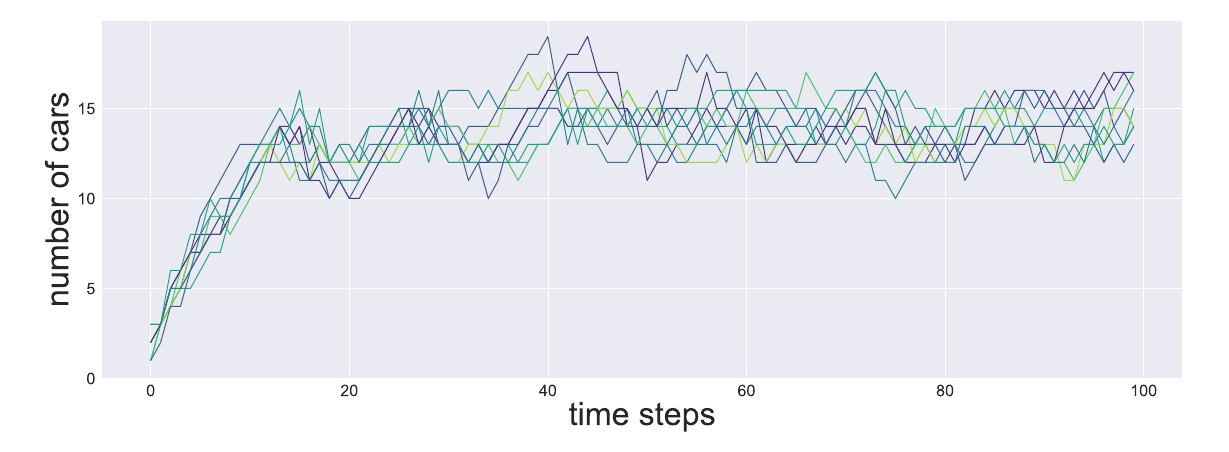}
\caption{The variation in the number of cars across ten different roads for 100 steps. Wherein each road has a different seed leading to different trends in the green/red signals in the exit cells, and a different number of cars are spawned at each step. Each road is seven cells long and three lanes wide, thus it can host a maximum of 21 cars.}\label{fig:ABMax_traffic_result}
\end{figure*}

A simulation of such a model at a smaller scale for 100 steps is shown in Figure~\ref{fig:ABMax_traffic_result}. 

\subsection{Financial market}
\begin{figure}[h]
\centering
\includegraphics[width=1.0\linewidth]{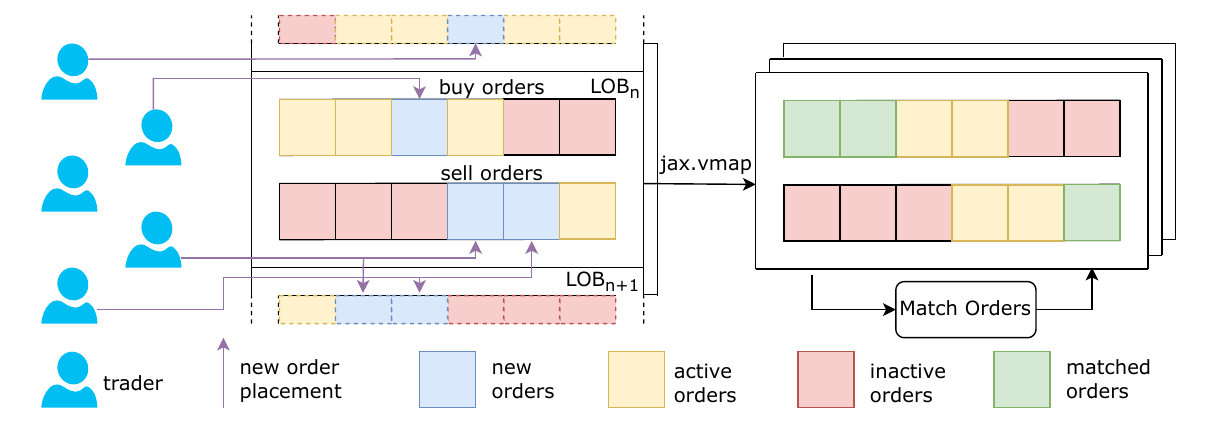}
\caption{Noisy traders place buy or sell orders across multiple limit order books (LOBs), where orders may be active, inactive, or new. After each step, when all traders have placed their orders, all LOBs are matched in parallel. Using ABMax data structures and JAX’s \texttt{vmap}, this matching is vectorized across every book. }\label{fig:ABMax_finance}
\end{figure}

In this model, we simulated a financial market of noisy traders interacting through a limit order book (LOB). Such simulations of LOB have been implemented before in JAX~\cite{frey2023jax}, which focused on matching one order in a step. Here, we developed a new matching algorithm where multiple orders can be matched in parallel. This step requires sorting the buy and sell orders according to price and calculating the cumulative sum of shares in the sorted list of orders. Vectorization of the order matching step allows simulation of multiple LOBs in parallel, as shown in Figure~\ref{fig:ABMax_finance}. Thus, leading to a simulation of a financial market where traders can invest in multiple stock markets. In this simulation, we modeled the traders and their orders as agents by inheriting the \texttt{Agent} class from ABMax.

Table~\ref{tab:finance_tab} again captures the gain in efficiency captured by using \texttt{vmap} across the number of LOBs, which were modelled as ABMax data structures.

\begin{table}[h]
\caption{Time taken in simulating 100 steps of finance markets with different numbers of limit order books (LOBs) of maximum size 1000 each.}
\label{tab:finance_tab}
\centering
\begin{tabular}{ll}
number of LOBs & time taken (ms)\\ \midrule
 $10$   & $297.01$\\ 
 $20$   & $333.47$\\
 $50$   & $470.88$\\
 $100$  & $663.80$\\
 $200$  & $1034.36$\\
 $500$  & $2212.47$\\
\end{tabular}
\end{table}

\begin{figure*}[h]
\centering
\includegraphics[width=1.0\linewidth]{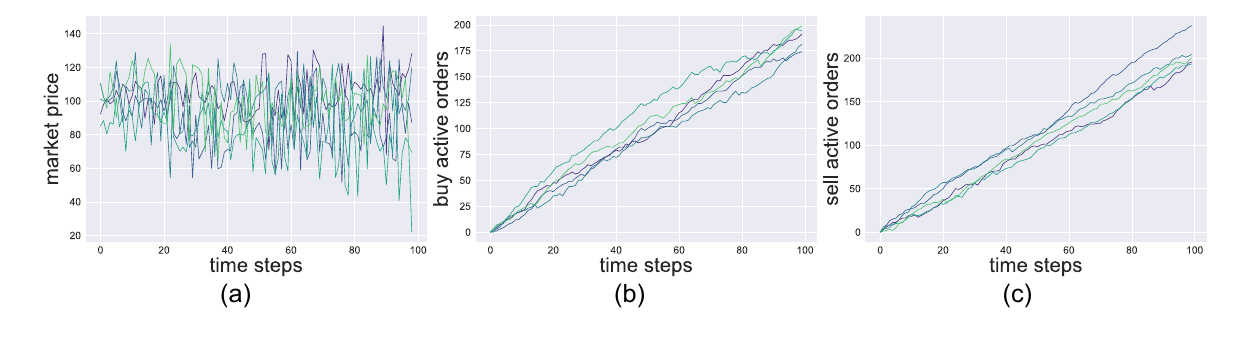}
\caption{The \textbf{(a)} market price, \textbf{(b)} number of active buy orders, and \textbf{(c)} number of active sell orders across five limit order books (LOBs) are simulated for 100 steps. The maximum size of each LOB is 1000.}\label{fig:lob_price_trend}
\end{figure*}

Figure~\ref{fig:lob_price_trend} illustrates the simulation states when ten noisy traders bid across five LOBs. In each LOB the price of the stock is initialized near the value of 100. The rapid divergence of stock prices based on different bidding patterns across different LOBs is evident in the results. 

\section{Discussion}

In this paper, we introduced a JAX-based ABM framework called ABMax. While doing so, we introduced two ways in which unique updates can be applied to a dynamically selected number of agents, in the form of the Rank-Match and the Sort-Count-Iterate algorithms. These algorithms are friendly to JAX transformations such that they can be \texttt{JIT}-compiled and can be run in parallel using JAX vectorization decorators like \texttt{vmap} and \texttt{pmap}. The chief motivation for doing this was to increase the efficiency of the serial-computation bottleneck in ABM when such simulations are run on hardware accelerators like GPUs. It is to be noted that these algorithms do not represent an exhaustive list of approaches to carry out this functionality, and future versions of ABMax may implement more efficient versions of the same.

Using ABMax, we also developed algorithms that are specific to the context of traffic modeling using cellular-automata and limit order book manipulation in a financial market. While developing them, one common theme that kept reappearing was that brute-force-based approaches, when vectorized, often outperformed an approach optimized for serial computations. This observation brings to light the fact that, with the rise in popularity of hardware accelerators, we might need to revisit how we formulate algorithms. A similar observation was  recently noted in the field of genetic-programming~\citep{de2025kozax}. The goal in such an exercise would be to investigate if a vectorized brute-force variant of the formulation exists and if it can, by virtue of vectorization, outperform the streamlined variant designed for serial computations. Through ABMax, we aim at maintaining a collection of such algorithms focused on ABM that are also friendly to JAX transforms.

Compared to other ABM libraries, ABMax does not aim to beat the state of the art in terms of performance or time taken per simulation step. For instance, Agents.jl~\citep{Agents.jl} and FLAME GPU2~\citep{richmond2023flame} represent the state of the art for ABM on CPU and hardware accelerators like GPU, respectively. The main aim of ABMax is to offer a middle-ground solution where certain serial computations related to ABM can be run efficiently using JIT-compilation and a one-liner automatic vectorization from JAX. This can be useful to reduce the amount of boilerplate code needed to run multiple agent-based models in parallel.

\balance

Finally, the aim of ABMax is also to maintain a diverse portfolio of agent-based models as examples and references. Right now, the examples include models of traffic jams and a financial market. In the future, we aim to include examples from power grids, inter-cropping, ecology, disease spread models, and weather forecasts. Such models, when scaled, can provide an ideal environment for the field of collective AI by providing adequate and realistic structures of interaction~\citep{epstein1996growing}.



\begin{acks}
This publication is part of the project Dutch Brain Interface Initiative (DBI$^2$) with project number 024.005.022 of the research programme Gravitation which is (partly) financed by the Dutch Research Council (NWO).
\end{acks}



\bibliographystyle{ACM-Reference-Format} 
\bibliography{sample}


\begin{thebibliography}{30}


\ifx \showCODEN    \undefined \def \showCODEN     #1{\unskip}     \fi
\ifx \showDOI      \undefined \def \showDOI       #1{#1}\fi
\ifx \showISBNx    \undefined \def \showISBNx     #1{\unskip}     \fi
\ifx \showISBNxiii \undefined \def \showISBNxiii  #1{\unskip}     \fi
\ifx \showISSN     \undefined \def \showISSN      #1{\unskip}     \fi
\ifx \showLCCN     \undefined \def \showLCCN      #1{\unskip}     \fi
\ifx \shownote     \undefined \def \shownote      #1{#1}          \fi
\ifx \showarticletitle \undefined \def \showarticletitle #1{#1}   \fi
\ifx \showURL      \undefined \def \showURL       {\relax}        \fi
\providecommand\bibfield[2]{#2}
\providecommand\bibinfo[2]{#2}
\providecommand\natexlab[1]{#1}
\providecommand\showeprint[2][]{arXiv:#2}

\bibitem[\protect\citeauthoryear{Bezanson, Karpinski, Shah, and Edelman}{Bezanson et~al\mbox{.}}{2012}]%
        {bezanson2012julia}
\bibfield{author}{\bibinfo{person}{Jeff Bezanson}, \bibinfo{person}{Stefan Karpinski}, \bibinfo{person}{Viral~B Shah}, {and} \bibinfo{person}{Alan Edelman}.} \bibinfo{year}{2012}\natexlab{}.
\newblock \showarticletitle{Julia: A fast dynamic language for technical computing}.
\newblock \bibinfo{journal}{\emph{ArXiv preprint arXiv:1209.5145}} (\bibinfo{year}{2012}).
\newblock


\bibitem[\protect\citeauthoryear{Bradbury, Frostig, Hawkins, Johnson, Leary, Maclaurin, Necula, Paszke, Vander{P}las, Wanderman-{M}ilne, and Zhang}{Bradbury et~al\mbox{.}}{2018}]%
        {jax2018github}
\bibfield{author}{\bibinfo{person}{James Bradbury}, \bibinfo{person}{Roy Frostig}, \bibinfo{person}{Peter Hawkins}, \bibinfo{person}{Matthew~James Johnson}, \bibinfo{person}{Chris Leary}, \bibinfo{person}{Dougal Maclaurin}, \bibinfo{person}{George Necula}, \bibinfo{person}{Adam Paszke}, \bibinfo{person}{Jake Vander{P}las}, \bibinfo{person}{Skye Wanderman-{M}ilne}, {and} \bibinfo{person}{Qiao Zhang}.} \bibinfo{year}{2018}\natexlab{}.
\newblock \bibinfo{title}{{JAX}: composable transformations of {P}ython+{N}um{P}y programs}.
\newblock
\newblock
\urldef\tempurl%
\url{http://github.com/jax-ml/jax}
\showURL{%
\tempurl}


\bibitem[\protect\citeauthoryear{Chen}{Chen}{2012}]%
        {chen2012agent}
\bibfield{author}{\bibinfo{person}{Liang Chen}.} \bibinfo{year}{2012}\natexlab{}.
\newblock \showarticletitle{Agent-based modeling in urban and architectural research: A brief literature review}.
\newblock \bibinfo{journal}{\emph{Frontiers Of Architectural Research}} \bibinfo{volume}{1}, \bibinfo{number}{2} (\bibinfo{year}{2012}), \bibinfo{pages}{166--177}.
\newblock


\bibitem[\protect\citeauthoryear{Datseris, Vahdati, and DuBois}{Datseris et~al\mbox{.}}{2022}]%
        {Agents.jl}
\bibfield{author}{\bibinfo{person}{George Datseris}, \bibinfo{person}{Ali~R. Vahdati}, {and} \bibinfo{person}{Timothy~C. DuBois}.} \bibinfo{year}{2022}\natexlab{}.
\newblock \showarticletitle{Agents.jl: a performant and feature-full agent-based modeling software of minimal code complexity}.
\newblock \bibinfo{journal}{\emph{{SIMULATION}}} \bibinfo{volume}{0}, \bibinfo{number}{0} (\bibinfo{date}{Jan.} \bibinfo{year}{2022}), \bibinfo{pages}{003754972110688}.
\newblock


\bibitem[\protect\citeauthoryear{De~Vries, Keemink, and van Gerven}{De~Vries et~al\mbox{.}}{2025}]%
        {de2025kozax}
\bibfield{author}{\bibinfo{person}{Sigur De~Vries}, \bibinfo{person}{Sander~Wessel Keemink}, {and} \bibinfo{person}{Marcel Antonius~Johannes van Gerven}.} \bibinfo{year}{2025}\natexlab{}.
\newblock \showarticletitle{Kozax: Flexible and Scalable Genetic Programming in JAX}. In \bibinfo{booktitle}{\emph{Proceedings of the Genetic And Evolutionary Computation Conference Companion}}. \bibinfo{pages}{603--606}.
\newblock


\bibitem[\protect\citeauthoryear{DeAngelis and Diaz}{DeAngelis and Diaz}{2019}]%
        {deangelis2019decision}
\bibfield{author}{\bibinfo{person}{Donald~L DeAngelis} {and} \bibinfo{person}{Stephanie~G Diaz}.} \bibinfo{year}{2019}\natexlab{}.
\newblock \showarticletitle{Decision-making in agent-based modeling: A current review and future prospectus}.
\newblock \bibinfo{journal}{\emph{Frontiers In Ecology And Evolution}}  \bibinfo{volume}{6} (\bibinfo{year}{2019}), \bibinfo{pages}{237}.
\newblock


\bibitem[\protect\citeauthoryear{Epstein and Axtell}{Epstein and Axtell}{1996}]%
        {epstein1996growing}
\bibfield{author}{\bibinfo{person}{Joshua~M Epstein} {and} \bibinfo{person}{Robert Axtell}.} \bibinfo{year}{1996}\natexlab{}.
\newblock \bibinfo{booktitle}{\emph{Growing artificial societies: social science from the bottom up}}.
\newblock \bibinfo{publisher}{Brookings Institution Press}.
\newblock


\bibitem[\protect\citeauthoryear{Fortuna and Troisi}{Fortuna and Troisi}{2010}]%
        {fortuna2010agent}
\bibfield{author}{\bibinfo{person}{Sara Fortuna} {and} \bibinfo{person}{Alessandro Troisi}.} \bibinfo{year}{2010}\natexlab{}.
\newblock \showarticletitle{Agent-based modeling for the 2D molecular self-organization of realistic molecules}.
\newblock \bibinfo{journal}{\emph{The Journal Of Physical Chemistry B}} \bibinfo{volume}{114}, \bibinfo{number}{31} (\bibinfo{year}{2010}), \bibinfo{pages}{10151--10159}.
\newblock


\bibitem[\protect\citeauthoryear{Foster}{Foster}{2005}]%
        {foster2005simplistic}
\bibfield{author}{\bibinfo{person}{John Foster}.} \bibinfo{year}{2005}\natexlab{}.
\newblock \showarticletitle{From simplistic to complex systems in economics}.
\newblock \bibinfo{journal}{\emph{Cambridge Journal Of Economics}} \bibinfo{volume}{29}, \bibinfo{number}{6} (\bibinfo{year}{2005}), \bibinfo{pages}{873--892}.
\newblock


\bibitem[\protect\citeauthoryear{Frey, Li, Nagy, Sapora, Lu, Zohren, Foerster, and Calinescu}{Frey et~al\mbox{.}}{2023}]%
        {frey2023jax}
\bibfield{author}{\bibinfo{person}{Sascha~Yves Frey}, \bibinfo{person}{Kang Li}, \bibinfo{person}{Peer Nagy}, \bibinfo{person}{Silvia Sapora}, \bibinfo{person}{Christopher Lu}, \bibinfo{person}{Stefan Zohren}, \bibinfo{person}{Jakob Foerster}, {and} \bibinfo{person}{Anisoara Calinescu}.} \bibinfo{year}{2023}\natexlab{}.
\newblock \showarticletitle{JAX-LOB: A GPU-Accelerated limit order book simulator to unlock large scale reinforcement learning for trading}. In \bibinfo{booktitle}{\emph{Proceedings of the Fourth ACM International Conference On AI In Finance}}. \bibinfo{pages}{583--591}.
\newblock


\bibitem[\protect\citeauthoryear{Heek, Levskaya, Oliver, Ritter, Rondepierre, Steiner, and van {Z}ee}{Heek et~al\mbox{.}}{2024}]%
        {flax2020github}
\bibfield{author}{\bibinfo{person}{Jonathan Heek}, \bibinfo{person}{Anselm Levskaya}, \bibinfo{person}{Avital Oliver}, \bibinfo{person}{Marvin Ritter}, \bibinfo{person}{Bertrand Rondepierre}, \bibinfo{person}{Andreas Steiner}, {and} \bibinfo{person}{Marc van {Z}ee}.} \bibinfo{year}{2024}\natexlab{}.
\newblock \bibinfo{booktitle}{\emph{{F}lax: A neural network library and ecosystem for {JAX}}}.
\newblock
\urldef\tempurl%
\url{http://github.com/google/flax}
\showURL{%
\tempurl}


\bibitem[\protect\citeauthoryear{Jiang, Sengupta, Demmel, and Williams}{Jiang et~al\mbox{.}}{2024}]%
        {jiang2024large}
\bibfield{author}{\bibinfo{person}{Xuan Jiang}, \bibinfo{person}{Raja Sengupta}, \bibinfo{person}{James Demmel}, {and} \bibinfo{person}{Samuel Williams}.} \bibinfo{year}{2024}\natexlab{}.
\newblock \showarticletitle{Large scale multi-GPU based parallel traffic simulation for accelerated traffic assignment and propagation}.
\newblock \bibinfo{journal}{\emph{Transportation Research Part C: Emerging Technologies}}  \bibinfo{volume}{169} (\bibinfo{year}{2024}), \bibinfo{pages}{104873}.
\newblock


\bibitem[\protect\citeauthoryear{Kazil, Masad, and Crooks}{Kazil et~al\mbox{.}}{2020}]%
        {python-mesa-2020}
\bibfield{author}{\bibinfo{person}{Jackie Kazil}, \bibinfo{person}{David Masad}, {and} \bibinfo{person}{Andrew Crooks}.} \bibinfo{year}{2020}\natexlab{}.
\newblock \showarticletitle{Utilizing Python for Agent-Based Modeling: The Mesa Framework}. In \bibinfo{booktitle}{\emph{International Conference on Social Computing, Behavioral-Cultural Modeling and Prediction and Behavior Representation in Modeling and Simulation}}. \bibinfo{publisher}{Springer International Publishing}, \bibinfo{address}{Cham}, \bibinfo{pages}{308--317}.
\newblock


\bibitem[\protect\citeauthoryear{Kulkarni and Bassett}{Kulkarni and Bassett}{2025}]%
        {kulkarni2024towards}
\bibfield{author}{\bibinfo{person}{Suman Kulkarni} {and} \bibinfo{person}{Dani~S Bassett}.} \bibinfo{year}{2025}\natexlab{}.
\newblock \showarticletitle{Towards principles of brain network organization and function}.
\newblock \bibinfo{journal}{\emph{{A}nnual {R}eview Of {B}iophysics}}  \bibinfo{volume}{54} (\bibinfo{year}{2025}).
\newblock


\bibitem[\protect\citeauthoryear{Kwapie{\'n} and Dro{\.z}d{\.z}}{Kwapie{\'n} and Dro{\.z}d{\.z}}{2012}]%
        {kwapien2012physical}
\bibfield{author}{\bibinfo{person}{Jaros{\l}aw Kwapie{\'n}} {and} \bibinfo{person}{Stanis{\l}aw Dro{\.z}d{\.z}}.} \bibinfo{year}{2012}\natexlab{}.
\newblock \showarticletitle{Physical approach to complex systems}.
\newblock \bibinfo{journal}{\emph{Physics Reports}} \bibinfo{volume}{515}, \bibinfo{number}{3-4} (\bibinfo{year}{2012}), \bibinfo{pages}{115--226}.
\newblock


\bibitem[\protect\citeauthoryear{Lange}{Lange}{2023}]%
        {evosax2022github}
\bibfield{author}{\bibinfo{person}{Robert~Tjarko Lange}.} \bibinfo{year}{2023}\natexlab{}.
\newblock \showarticletitle{evosax: JAX-Based Evolution Strategies}. In \bibinfo{booktitle}{\emph{Proceedings of the Companion Conference On Genetic And Evolutionary Computation}} (Lisbon, Portugal) \emph{(\bibinfo{series}{GECCO '23 Companion})}. \bibinfo{publisher}{Association for Computing Machinery}, \bibinfo{address}{New York, NY, USA}, \bibinfo{pages}{659–662}.
\newblock
\showISBNx{9798400701207}


\bibitem[\protect\citeauthoryear{Lu, Kuba, Letcher, Metz, Schroeder~de Witt, and Foerster}{Lu et~al\mbox{.}}{2022}]%
        {lu2022discovered}
\bibfield{author}{\bibinfo{person}{Chris Lu}, \bibinfo{person}{Jakub Kuba}, \bibinfo{person}{Alistair Letcher}, \bibinfo{person}{Luke Metz}, \bibinfo{person}{Christian Schroeder~de Witt}, {and} \bibinfo{person}{Jakob Foerster}.} \bibinfo{year}{2022}\natexlab{}.
\newblock \showarticletitle{Discovered policy optimisation}.
\newblock \bibinfo{journal}{\emph{Advances In Neural Information Processing Systems}}  \bibinfo{volume}{35} (\bibinfo{year}{2022}), \bibinfo{pages}{16455--16468}.
\newblock


\bibitem[\protect\citeauthoryear{Macy and Willer}{Macy and Willer}{2002}]%
        {macy2002factors}
\bibfield{author}{\bibinfo{person}{Michael~W Macy} {and} \bibinfo{person}{Robert Willer}.} \bibinfo{year}{2002}\natexlab{}.
\newblock \showarticletitle{From factors to actors: Computational sociology and agent-based modeling}.
\newblock \bibinfo{journal}{\emph{Annual Review Of Sociology}} \bibinfo{volume}{28}, \bibinfo{number}{1} (\bibinfo{year}{2002}), \bibinfo{pages}{143--166}.
\newblock


\bibitem[\protect\citeauthoryear{Ohsawa, Akuzawa, Matsushima, Bezerra, Iwasawa, Kajino, Takenaka, and Matsuo}{Ohsawa et~al\mbox{.}}{2018}]%
        {ohsawa2018neuron}
\bibfield{author}{\bibinfo{person}{Shohei Ohsawa}, \bibinfo{person}{Kei Akuzawa}, \bibinfo{person}{Tatsuya Matsushima}, \bibinfo{person}{Gustavo Bezerra}, \bibinfo{person}{Yusuke Iwasawa}, \bibinfo{person}{Hiroshi Kajino}, \bibinfo{person}{Seiya Takenaka}, {and} \bibinfo{person}{Yutaka Matsuo}.} \bibinfo{year}{2018}\natexlab{}.
\newblock \showarticletitle{Neuron as an Agent}.
\newblock  (\bibinfo{year}{2018}).
\newblock


\bibitem[\protect\citeauthoryear{Park}{Park}{2005}]%
        {park2005internet}
\bibfield{author}{\bibinfo{person}{Kihong Park}.} \bibinfo{year}{2005}\natexlab{}.
\newblock \bibinfo{title}{The Internet as a Complex System}.
\newblock
\newblock


\bibitem[\protect\citeauthoryear{Pio-Lopez, Bischof, LaPalme, and Levin}{Pio-Lopez et~al\mbox{.}}{2023}]%
        {pio2022scaling}
\bibfield{author}{\bibinfo{person}{Leo Pio-Lopez}, \bibinfo{person}{Johanna Bischof}, \bibinfo{person}{Jennifer~V LaPalme}, {and} \bibinfo{person}{Michael Levin}.} \bibinfo{year}{2023}\natexlab{}.
\newblock \showarticletitle{The scaling of goals via homeostasis: an evolutionary simulation, experiment and analysis}.
\newblock \bibinfo{journal}{\emph{Interface Focus}} \bibinfo{volume}{13}, \bibinfo{number}{3} (\bibinfo{year}{2023}).
\newblock


\bibitem[\protect\citeauthoryear{Richmond and Chimeh}{Richmond and Chimeh}{2017}]%
        {richmond2017flame}
\bibfield{author}{\bibinfo{person}{Paul Richmond} {and} \bibinfo{person}{Mozhgan~K Chimeh}.} \bibinfo{year}{2017}\natexlab{}.
\newblock \showarticletitle{Flame gpu: Complex system simulation framework}. In \bibinfo{booktitle}{\emph{2017 International Conference on High Performance Computing \& Simulation (HPCS)}}. IEEE, \bibinfo{pages}{11--17}.
\newblock


\bibitem[\protect\citeauthoryear{Richmond, Chisholm, Heywood, Chimeh, and Leach}{Richmond et~al\mbox{.}}{2023}]%
        {richmond2023flame}
\bibfield{author}{\bibinfo{person}{Paul Richmond}, \bibinfo{person}{Robert Chisholm}, \bibinfo{person}{Peter Heywood}, \bibinfo{person}{Mozhgan~Kabiri Chimeh}, {and} \bibinfo{person}{Matthew Leach}.} \bibinfo{year}{2023}\natexlab{}.
\newblock \showarticletitle{FLAME GPU 2: A framework for flexible and performant agent based simulation on GPUs}.
\newblock \bibinfo{journal}{\emph{Software: Practice And Experience}} \bibinfo{volume}{53}, \bibinfo{number}{8} (\bibinfo{year}{2023}), \bibinfo{pages}{1659--1680}.
\newblock


\bibitem[\protect\citeauthoryear{Rutter, Savona, Glonti, Bibby, Cummins, Finegood, Greaves, Harper, Hawe, Moore, et~al\mbox{.}}{Rutter et~al\mbox{.}}{2017}]%
        {rutter2017need}
\bibfield{author}{\bibinfo{person}{Harry Rutter}, \bibinfo{person}{Natalie Savona}, \bibinfo{person}{Ketevan Glonti}, \bibinfo{person}{Jo Bibby}, \bibinfo{person}{Steven Cummins}, \bibinfo{person}{Diane~T Finegood}, \bibinfo{person}{Felix Greaves}, \bibinfo{person}{Laura Harper}, \bibinfo{person}{Penelope Hawe}, \bibinfo{person}{Laurence Moore}, {et~al\mbox{.}}} \bibinfo{year}{2017}\natexlab{}.
\newblock \showarticletitle{The need for a complex systems model of evidence for public health}.
\newblock \bibinfo{journal}{\emph{The Lancet}} \bibinfo{volume}{390}, \bibinfo{number}{10112} (\bibinfo{year}{2017}), \bibinfo{pages}{2602--2604}.
\newblock


\bibitem[\protect\citeauthoryear{Samanidou, Zschischang, Stauffer, and Lux}{Samanidou et~al\mbox{.}}{2007}]%
        {samanidou2007agent}
\bibfield{author}{\bibinfo{person}{Egle Samanidou}, \bibinfo{person}{Elmar Zschischang}, \bibinfo{person}{Dietrich Stauffer}, {and} \bibinfo{person}{Thomas Lux}.} \bibinfo{year}{2007}\natexlab{}.
\newblock \showarticletitle{Agent-based models of financial markets}.
\newblock \bibinfo{journal}{\emph{Reports On Progress In Physics}} \bibinfo{volume}{70}, \bibinfo{number}{3} (\bibinfo{year}{2007}), \bibinfo{pages}{409}.
\newblock


\bibitem[\protect\citeauthoryear{Testa and Bojarski}{Testa and Bojarski}{2000}]%
        {testa2000molecules}
\bibfield{author}{\bibinfo{person}{Bernard Testa} {and} \bibinfo{person}{Andrzej~J Bojarski}.} \bibinfo{year}{2000}\natexlab{}.
\newblock \showarticletitle{Molecules as complex adaptative systems: Constrained molecular properties and their biochemical significance}.
\newblock \bibinfo{journal}{\emph{European Journal Of Pharmaceutical Sciences}}  \bibinfo{volume}{11} (\bibinfo{year}{2000}), \bibinfo{pages}{S3--S14}.
\newblock


\bibitem[\protect\citeauthoryear{Wilensky}{Wilensky}{1997}]%
        {Wilensky1997}
\bibfield{author}{\bibinfo{person}{Uri Wilensky}.} \bibinfo{year}{1997}\natexlab{}.
\newblock \bibinfo{title}{NetLogo Wolf Sheep Predation model}.
\newblock
\newblock
\urldef\tempurl%
\url{http://ccl.northwestern.edu/netlogo/models/WolfSheepPredation}
\showURL{%
\tempurl}


\bibitem[\protect\citeauthoryear{Wilensky}{Wilensky}{1998}]%
        {wilensky1998netlogo_flocking}
\bibfield{author}{\bibinfo{person}{Uri Wilensky}.} \bibinfo{year}{1998}\natexlab{}.
\newblock \bibinfo{title}{{NetLogo} Flocking Model}.
\newblock \bibinfo{howpublished}{\url{http://ccl.northwestern.edu/netlogo/models/Flocking}}.
\newblock


\bibitem[\protect\citeauthoryear{Xiao, Andelfinger, Eckhoff, Cai, and Knoll}{Xiao et~al\mbox{.}}{2019}]%
        {xiao2019survey}
\bibfield{author}{\bibinfo{person}{Jiajian Xiao}, \bibinfo{person}{Philipp Andelfinger}, \bibinfo{person}{David Eckhoff}, \bibinfo{person}{Wentong Cai}, {and} \bibinfo{person}{Alois Knoll}.} \bibinfo{year}{2019}\natexlab{}.
\newblock \showarticletitle{A survey on agent-based simulation using hardware accelerators}.
\newblock \bibinfo{journal}{\emph{ACM Computing Surveys (CSUR)}} \bibinfo{volume}{51}, \bibinfo{number}{6} (\bibinfo{year}{2019}), \bibinfo{pages}{1--35}.
\newblock


\bibitem[\protect\citeauthoryear{Yang, Andelfinger, Cai, and Knoll}{Yang et~al\mbox{.}}{2018}]%
        {yang2018evaluation}
\bibfield{author}{\bibinfo{person}{Mingyu Yang}, \bibinfo{person}{Philipp Andelfinger}, \bibinfo{person}{Wentong Cai}, {and} \bibinfo{person}{Alois Knoll}.} \bibinfo{year}{2018}\natexlab{}.
\newblock \showarticletitle{Evaluation of Conflict Resolution Methods for Agent-Based Simulations on the GPU}. In \bibinfo{booktitle}{\emph{Proceedings of the 2018 ACM SIGSIM Conference On Principles Of Advanced Discrete Simulation}}. \bibinfo{pages}{129--132}.
\newblock


\end{thebibliography}


\end{document}